%% file: main.tex
\documentclass[titlepage,10pt]{article}

\usepackage[letterpaper,top=2cm,bottom=2cm,left=3cm,right=3cm,marginparwidth=1.75cm]{geometry}

\usepackage{graphicx} 
\usepackage{caption}
\usepackage{subcaption}

\usepackage{csquotes}
\usepackage{enumitem}
\usepackage{xcolor}
\usepackage{amsmath,amssymb}
\usepackage{orcidlink}                
\usepackage[affil-it,blocks]{authblk} 

\begin{document}
\title{Letter of Intent for \\Alpha Knockout from Nuclei in Hard Exclusive Processes with ALERT\\(PAC54)}

\include{authors} 

\maketitle

\section*{Executive Summary}

We propose to measure hard exclusive $\alpha$-knockout reactions from the
light $\alpha$-conjugate nuclei $^{12}$C, $^{16}$O, and $^{20}$Ne using
electron scattering. Three reaction channels will be measured for the first
time on the same target set: (i) the quasi-elastic knockout $(e,e^\prime\alpha)$
knockout, which establishes a baseline against $\alpha$-clustering
model calculations and tests final-state-interaction (FSI) models; (ii) coherent
deeply virtual Compton scattering (DVCS) on a bound $\alpha$, which provides
a quark-level probe of the bound system through the beam-spin asymmetry; and
(iii) coherent vector-meson production ($\phi$ and J/$\psi$) on a bound
$\alpha$, which probes the gluon distribution of the bound cluster through
two-gluon exchange. The central observables are ratios of cross sections and
beam-spin asymmetries between the bound $\alpha$ in $^{12}$C, $^{16}$O, and
$^{20}$Ne and the free $^4$He ground state. Together these measurements
directly address the question of \emph{the role of QCD in the structure of
	light nuclei}: do the quark and gluon distributions of an $\alpha$ embedded
in a heavier nucleus differ from those of a free $^4$He, and do those
modifications track the increasing $\alpha$-clustering strength along the
$\alpha$-conjugate chain~\cite{Tohsaki:2001an,Freer:2017gip,Otsuka:2022bcf,Harris:2025alpha}?

The measurement will use the 12~GeV CEBAF accelerator~\cite{Adderley:2024czm} in
Hall~B with the CLAS12 spectrometer instrumented with the ALERT low-energy
recoil detector and high-pressure gas straw targets. ALERT, which replaced
the CLAS12 silicon vertex tracker, was successfully commissioned and operated
during Run Group L (RG-L) from April through August~2025. It uniquely
provides identification of low-momentum recoiling light ions
($p,\,d,\,t,\,^3$He, and $^4$He) close to the beamline at the high
luminosities required by these exclusive channels. The experiment is a
natural extension of the ALERT physics program (RG-L tagged-EMC, SRC, and
coherent processes on $^4$He) and of the recently conditionally approved
J/$\psi$-on-$^4$He measurement (PAC53 C2, 100~days~\cite{ALERT:Cprime}).
Matching the accumulated charge for the helium target during RG-L,
we estimate a total of \textbf{85 PAC days}, including short low-energy
($2.2$~GeV) calibration runs on each gas target and 20-day high-energy
($10.6$~GeV) production runs on $^4$He, $^{16}$O, $^{20}$Ne, and ethane
(C$_2$H$_6$, providing the $^{12}$C target).

\newpage

\tableofcontents

\newpage

\section{Introduction}

The $\alpha$ particle has occupied a central place in nuclear physics since
its inception. Already in the first textbook on nuclear physics, Rutherford,
Chadwick, and Ellis (1930) wrote\footnote{This book was published about 2 year before the neutron was discovered.}
: \emph{``The general evidence on nuclei
	strongly supports the view that the $\alpha$ particle is of primary
	importance as a unit of the structure of nuclei in general and particularly
	of the heavier elements. It seems very possible that the greater part of the
	mass of heavy nuclei is due to $\alpha$ particles which have an independent
	existence in the nuclear structure.''} Nearly a century later, $\alpha$
clustering remains a defining feature of the structure of light
nuclei~\cite{Freer:2017gip}, while the description of nuclear structure has
remained almost entirely \emph{hadronic}, in terms of nucleon (or nucleonic
cluster) degrees of freedom. The fundamental theory underlying the strong
interaction, quantum chromodynamics (QCD), enters the description only
through the static properties of the nucleons themselves. The central
question motivating this proposal is whether QCD plays a more direct role in
the structure of light nuclei---and, in particular, whether the quark and
gluon structure of an $\alpha$ \emph{cluster} bound inside a larger nucleus
is modified relative to that of a free $^4$He.

\begin{figure}[h!]
	\centering
	\includegraphics[width=0.5\textwidth]{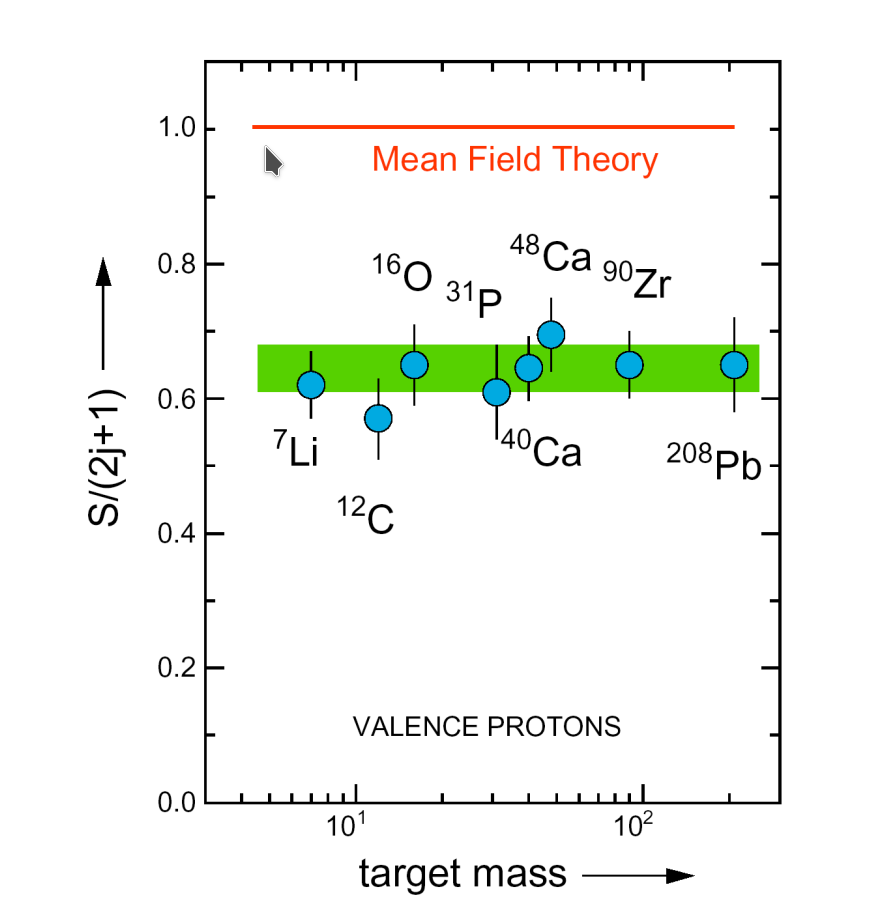}
	\caption{Single-particle spectroscopic factors $S$ extracted from
		quasi-elastic $(e,e^\prime p)$ across nuclei. The systematic
		reduction to $S\simeq 0.65$ relative to the independent-particle
		limit $S=1$ has stood as an open puzzle for decades.
		Reproduced from~\cite{Aumann:2020tcq,Lapikas:1993uwd}.
		\label{fig:SvsA}}
\end{figure}

\begin{figure}[h!]
	\centering
	\includegraphics[width=0.5\textwidth]{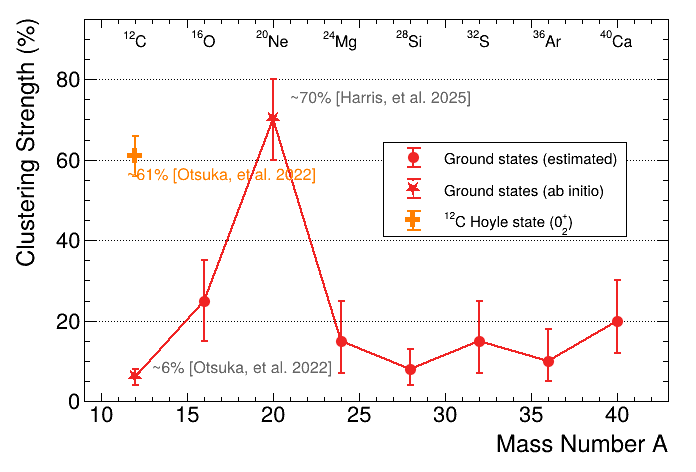}
	\caption{Ground-state $\alpha$-clustering strength as a function of mass
		number for light $\alpha$-conjugate nuclei. The strength rises from
		$\sim 6\%$ in $^{12}$C to $\sim 70\%$ in $^{20}$Ne and saturates
		near the cluster-model maximum in the Hoyle state of
		$^{12}$C~\cite{Tohsaki:2001an,Otsuka:2022bcf,Harris:2025alpha}.
		Values for $^{12}$C and the Hoyle state are the ab initio
		cluster-decomposition probabilities of
		Ref.~\cite{Otsuka:2022bcf}; the $^{16}$O and $^{20}$Ne points
		are taken from the model-independent ANC analysis of
		Ref.~\cite{Harris:2025alpha} and the SA-NCSM ab initio
		calculation of Ref.~\cite{Dreyfuss:2020yhq}; entries for
		$^{24}$Mg, $^{28}$Si, $^{32}$S, and $^{36}$Ar are estimated
		from the shell-model $\alpha$-spectroscopic factors of
		Ref.~\cite{Chung:1978zz} and the self-consistent mean-field
		calculations of Ref.~\cite{Maruhn:2006pf}; the $^{40}$Ca point
		uses the AMD+GCM result of Ref.~\cite{Taniguchi:2007zz}.
		\label{fig:clusterStrength}}
\end{figure}

Three observations frame this question. First, the EMC effect demonstrates
that nucleon parton distributions are modified inside nuclei, and the effect
is already nearly at full strength in $^4$He~\cite{Seely:2009gt}, despite its
small size and low average density. The origin of the EMC effect remains
unresolved more than four decades after its discovery. Second, the
nucleon spectroscopic factors extracted from quasi-elastic $(e,e^\prime p)$
scattering show a long-standing and systematic quenching of the
single-particle strength relative to the independent-particle limit,
\begin{equation}
	S_{nlj} = \sum_{m,M_A} \big| \langle \Psi_{A+1,J_f M_f} | a^{\dagger}_{nljm} |
	\Psi_{A,J_i M_i}\rangle \big|^2,
	\label{eq:Snlj}
\end{equation}
typically yielding $S\simeq 0.65$ across a wide range of
nuclei~\cite{Lapikas:1993uwd,Aumann:2020tcq} (see
Fig.~\ref{fig:SvsA}). Short-range correlations (SRC), although qualitatively
important, account for only $10$--$15\%$ of the missing
strength~\cite{Hen:2016kwk}, leaving the dominant remainder to be supplied by
long-range correlations or by other mechanisms not present in the
independent-particle picture~\cite{Dickhoff:2004xx,Aumann:2020tcq}. Third,
$\alpha$ clustering is itself a long-range correlation: ab initio
calculations and modern $\alpha$-transfer measurements quantify the
clustering strength of the ground states of $^{12}$C, $^{16}$O, and $^{20}$Ne
as $\sim 6\%$, $\sim 30\%$, and $\sim 70\%$ of the maximal cluster-model
value, respectively (Fig.~\ref{fig:clusterStrength}), with the Hoyle state of
$^{12}$C reaching $\sim 61\%$~\cite{Tohsaki:2001an,Otsuka:2022bcf,Harris:2025alpha}.
The growth of $\alpha$ clustering across this chain provides a controlled
laboratory in which to ask whether QCD-level (parton) modifications track
\emph{cluster} structure rather than nucleon structure.

The ALERT detector, which was installed in CLAS12 for the successful running of RG-L in 2025,
uniquely provides for the first time the experimental capability to perform
such hard exclusive measurements with low-energy recoil $\alpha$ detection. We
propose to use this capability to extend the ongoing ALERT program on the
partonic structure of $^4$He to the partonic structure of an $\alpha$
\emph{embedded} in light $\alpha$-conjugate nuclei.

\section{Physics Motivation}

\subsection{The role of QCD in the structure of $^4$He}

The $\alpha$ particle ($^4$He) sits at a unique intersection in nuclear
physics. It is the lightest doubly-closed-shell nucleus, the smallest
isoscalar nucleus, and the natural building block of $\alpha$-cluster
models. Its charge radius, $r_c = 1.676 \pm 0.008$~fm, is smaller than that
of the much lighter $A=3$ isobars ($^3$H: $1.755$~fm, $^3$He:
$1.959$~fm)~\cite{Camsonne:2014vfa}, indicative of an unusually tightly bound
system. The charge form factor exhibits a clear diffractive
minimum~\cite{Camsonne:2014vfa}, which in two-dimensional Fourier space
encodes the transverse charge profile.

Two features make $^4$He the central testing ground for QCD effects in
nuclei. First, the EMC effect on $^4$He has nearly the full strength
observed in much heavier nuclei~\cite{Seely:2009gt}, demonstrating that
substantial nuclear modification of the parton distributions is already
present in the lightest closed-shell nucleus. Second, the structure of
$^4$He admits non-nucleonic configurations, such as the Brodsky--West
hexadiquark \cite{West:2020rlk} and other hidden-color components, that have no analog in a
purely nucleonic
description~\cite{Brodsky:1983vf,Bashkanov:2013cla}. The ALERT physics
program addresses both features through a comprehensive set of measurements
on $^4$He: tagged DIS for the EMC effect, coherent DVCS for the
two-dimensional quark transverse profile~\cite{CLAS:2021ovm}, incoherent
DVCS for medium-modified bound nucleons~\cite{Hattawy:2018liu}, and coherent
J/$\psi$ and $\phi$ production for the gluon transverse
profile~\cite{ALERT:Cprime}. The pairing of coherent DVCS (charge/quark
profile) with coherent vector-meson production (gluon profile) is what
allows direct comparison of the transverse \emph{quark} and \emph{gluon}
distributions of the same target. Determining whether the gluon transverse
distribution of $^4$He follows the same density profile as the charge form
factor---and whether it possesses its own diffractive minimum and where it
sits in $|t|$---is one of the principal goals of the ALERT program and synergetic with 
the bound-$\alpha$ measurements proposed here.

\subsection{From the partonic structure of free $^4$He to bound $\alpha$s}

If the partonic structure of a free $^4$He is qualitatively understood, the
next question is whether the same partonic structure is preserved when the
$\alpha$ is embedded in a heavier nucleus. By analogy to the conventional
EMC effect, we ask: \emph{is there an ``$\alpha$-EMC effect''?} That is, do
the quark and gluon distributions of an $\alpha$ change when it is bound
inside $^{12}$C, $^{16}$O, or $^{20}$Ne, and how does the modification
depend on the cluster structure of the host nucleus? The proposed exclusive
$\alpha$-knockout reactions provide three independent probes of this
question:
\begin{itemize}
	\item \textbf{Quasi-elastic $(e,e^\prime\alpha)$:} establishes a
	      \emph{nucleonic} (or, more precisely, cluster-level) baseline for the
	      knockout cross section. Comparison to DWIA calculations using
	      $\alpha$-cluster wave functions from AMD, FMD, RGM, GCM, the algebraic
	      cluster model, and the supersolidity description provides a direct test of
	      the cluster picture and of FSI
	      models~\cite{KanadaEnyo:2012aa,Chernykh:2007zz,Bijker:2014tka,Ohkubo:2020rsu,DeMeyer:2001uv}.
	      The principal observable is the ratio of the $(e,e^\prime\alpha)$ cross
	      section per bound $\alpha$ to the elastic $e\,^4$He cross section as a
	      function of the missing momentum.
	\item \textbf{Coherent DVCS on a bound $\alpha$:} provides a
	      \emph{quark-level} probe of the bound system. At leading twist a spin-0
	      target has only one chiral-even GPD, so the formalism developed for
	      coherent DVCS on a free $^4$He~\cite{CLAS:2021ovm} carries over directly to
	      the bound case. The principal observable is the ratio of the bound-$\alpha$
	      beam-spin asymmetry (BSA) to that measured for free $^4$He in the same
	      kinematics, after accounting for the $\alpha$ momentum-distribution and
	      DWIA-level FSIs.
	\item \textbf{Coherent vector-meson production ($\phi$ and J/$\psi$) on a
		      bound $\alpha$:} provides a \emph{gluon-level} probe at leading order through two-gluon
	      exchange. The heavy charm-quark mass in J/$\psi$ production ensures that
	      gluon exchange dominates the
	      amplitude. The principal observable is the ratio of the bound-$\alpha$
	      $|t|$-differential cross section to that on free $^4$He, including any
	      shift in the diffractive minimum (if such a minimum is established by the
	      companion $^4$He measurement~\cite{ALERT:Cprime}).
\end{itemize}
The pattern of these three ratios across $^{12}$C, $^{16}$O, and $^{20}$Ne
reflects the increasing $\alpha$-clustering strength of
Fig.~\ref{fig:clusterStrength}. If the partonic modification of the bound
$\alpha$ is driven by cluster degrees of freedom, the ratios should track
the clustering strength; if the modification is driven by mean-field or
nucleon-level mechanisms, no such tracking should appear. Either outcome is
diagnostic and unique to this measurement.

\subsection{Connection to the structure of light nuclei}

The structure of $^{12}$C, $^{16}$O, and $^{20}$Ne is described by a rich
ecosystem of theoretical approaches: the nuclear shell model and its
hybridization with cluster degrees of freedom in the Cluster Shell Model
(CSM); microscopic cluster models (RGM and GCM); molecular-dynamics
approaches (AMD and FMD), which describe both the Hoyle state of $^{12}$C
and the tetrahedral $4\alpha$ ground state of $^{16}$O without presupposing
clustering~\cite{Chernykh:2007zz,KanadaEnyo:2012aa}; the algebraic cluster
model with explicit geometric symmetries ($D_{3h}$ for $^{12}$C, $T_d$ for
$^{16}$O)~\cite{Bijker:2014tka}; and the supersolidity description, in which
the $\alpha$-cluster system simultaneously displays crystalline and
condensate character~\cite{Ohkubo:2020rsu,Carlson:2014vla,Karliner:2016zzc}. These
approaches are largely complementary: shell-model and cluster pictures
co-exist, and the question of which set of degrees of freedom is most
economical depends on the state and the observable.

The supersolidity description in particular is suggestive: it requires a
``crystalline superfluid,'' a description that is paradoxical within a
purely hadronic picture but acquires a natural interpretation if the
condensate is identified with the gluon field and the Pauli blocking with
the underlying quark structure. Taken together, the observation that the
EMC effect saturates already in $^4$He~\cite{Seely:2009gt}, the long-standing
quenching of the single-particle spectroscopic factors that cannot be
explained by SRC alone~\cite{Hen:2016kwk,Aumann:2020tcq}, and the dominance
of $\alpha$ clustering in the ground states of light $\alpha$-conjugate
nuclei together motivate a direct, partonic-level probe of the bound
$\alpha$. The proposed measurements provide exactly such a probe.

\subsection{Observations of nuclear dependence to partonic structure}

To illustrate how these three ratios will be interpreted, consider the 
following hypothetical outcome. Suppose the measured ratios show:
\begin{itemize}
	\item the quasi-elastic $(e,e^\prime\alpha)$ and coherent DVCS ratios are 
	      approximately constant across $^{12}$C, $^{16}$O, and $^{20}$Ne, with no 
	      kinematic dependence;
	\item the coherent DVMP ratios track the $\alpha$-clustering strength 
	      predictions of Fig.~\ref{fig:clusterStrength}, also with no kinematic 
	      dependence.
\end{itemize}
In this scenario, the cluster wave function and the bound-$\alpha$ quark 
distribution show no detectable nuclear dependence, while the bound-$\alpha$ 
gluon distribution exhibits a non-trivial nuclear dependence correlated with 
clustering strength. 
We would conclude that probing the system with gluons is more readily 
exciting cluster structures than the case where the quark structure is probed.

Again, the above paragraph is a speculative example meant to demonstrate how 
the proposed measurements may yield exciting new insights into the role of QCD 
in light nuclei. Every possible scenario where there is a difference in the 
measured ratio between any of the processes will be an exciting result as it 
would indicate a difference in the structure of the light nuclei at the 
partonic level. The pattern of differences between the three ratios --- 
quasi-elastic, DVCS, and DVMP --- discriminates between cluster-driven, 
quark-driven, and gluon-driven mechanisms for the partonic-level modification in nuclei. 
Put another way, the scenario where the ratios from all three processes have the same 
dependence on the target nucleus would indicate, for the measured 
kinematics and level of precision, the partonic structures beyond the extent of  
$^4$He play no role in the structure of light nuclei --- an important experimental result nonetheless.

\paragraph{A Challenge for Theory.} New theory and model predictions for the 
experimental results from the different process, particularly those probing the 
partonic structure, provide an excellent opportunity to push forward our 
understanding of QCD in nuclear matter and its role in the structure of light 
nuclei. We will work with our theory colleagues  and encourage them to put their 
understanding of QCD to the test by developing new predictions for the 
observables outlined for this proposed measurement.

\section{Experimental Details}

The experiment will use the 12~GeV CEBAF accelerator~\cite{Adderley:2024czm} to deliver polarized beam to CLAS12 in Hall~B.

\subsection{Required Equipment}


The proposed measurements use the existing CLAS12 spectrometer in Hall~B
together with the ALERT low-energy recoil detector and high-pressure gas
straw targets. CLAS12 detects the scattered electron and the forward
high-momentum hadrons (the real photon for DVCS and the decay products of
the produced vector mesons for $\phi$ and J/$\psi$ production). ALERT,
installed in place of the CLAS12 silicon vertex tracker, detects the
recoiling $\alpha$ and the residual ions emerging from the target.

ALERT consists of two subsystems:
\begin{itemize}
	\item The \textbf{ALERT Hyperbolic Drift Chamber (AHDC)}, built at IJCLab
	      (Orsay), provides charged-particle tracking with $2$~mm wire spacing and a
	      $\pm 4^\circ$ hyperbolic stereo angle. Beam tests at the ALTO facility with
	      $6$--$18$~MeV protons and $\alpha$ beams demonstrated a track resolution of
	      $\sim 120\,\mu$m and detection efficiencies of $95\%$ (protons) and $99\%$
	      ($\alpha$). The AHDC is read out by DREAM front-end electronics.
	\item The \textbf{ALERT Time-of-Flight (ATOF) hodoscope}, built at Argonne,
	      is a barrel of scintillator bars and wedges (inner-bar thickness $3$~mm,
	      outer-wedge thickness $2$~cm) read out by SiPMs. Module tests with a
	      $^{241}$Am source demonstrated a single-channel time resolution of
	      $\sigma\simeq 85$--$125$~ps, comfortably meeting the $<150$~ps design
	      requirement. Combined drift-chamber tracking and TOF identifies the light
	      ions $p$, $d$, $t$, $^3$He, and $^4$He; the $^4$He and $d$ species, which
	      have the same $m/q$, are separated by combining momentum and energy-loss
	      information.
\end{itemize}
ALERT was commissioned and operated successfully in Hall~B during RG-L from
April through August~2025, with only one broken AHDC wire and one ATOF
module replacement during the entire run, demonstrating the maturity and
operational stability of the detector.

\subsection{Targets}

We will use the same high-pressure gas straw target hardware that was
operated during RG-L, managed by the JLab Hall-B detector support
group. During RG-L commissioning the target ran at a pressure of $65$~psig
with $^1$H, $^2$H, and $^4$He gases. For the present proposal the same
target geometry will be used with $^4$He, $^{16}$O$_2$, $^{20}$Ne, and
ethane (C$_2$H$_6$). Ethane is chosen as the practical $^{12}$C target:
it is a gas at the operating pressure and provides the carbon density
required for the program while keeping radiation length and beam-induced
backgrounds compatible with ALERT and CLAS12. The hydrogen content of the
ethane target additionally provides an in-situ luminosity normalization
through elastic $ep$ scattering.

\paragraph{Luminosity limit.} During RG-L the luminosity was set by the
tolerance of the CLAS12 forward detectors, balancing detector lifetime
against integrated luminosity. We adopt the same limit here and scale the
beam current per target species to maintain the same nucleon-equivalent
RG-L level luminosity. The resulting beam currents are tabulated in
Table~\ref{tab:beamtime}.

\subsection{Proposed Measurements}
\label{sec:proposed}

The proposed program consists of three exclusive channels measured on each
of the four targets ($^4$He, $^{12}$C, $^{16}$O, $^{20}$Ne) at $10.6$~GeV
beam energy:
\begin{enumerate}
	\item \textbf{Quasi-elastic $\alpha$ knockout, $A(e,e^\prime\alpha)X$.}
	      The principal observable is the ratio of the $(e,e^\prime\alpha)$ cross
	      section per bound $\alpha$ to the elastic $e\,^4$He cross section, in bins
	      of missing momentum and energy. This channel establishes the
	      cluster-level baseline against $\alpha$-clustering models, tests FSI
	      descriptions (Coulomb plus optical-model
	      distortions)~\cite{DeMeyer:2001uv}, and benchmarks the ALERT $\alpha$-tagging
	      performance for the harder exclusive channels below.

	\item \textbf{Coherent DVCS on a bound $\alpha$,
		      $A(e,e^\prime\alpha\gamma)X$.}
	      The principal observable is the beam-spin asymmetry as a function of the
	      azimuthal angle $\phi$ between the leptonic and hadronic planes, in bins of
	      $Q^2$, $x_B$, and $|t|$. The bound-$\alpha$ BSA is compared as a ratio to
	      the BSA measured for free $^4$He~\cite{CLAS:2021ovm} in the same kinematic
	      bins. Departures from unity probe the modification of the leading-twist
	      chiral-even GPD of the $\alpha$ in the nuclear medium.

	\item \textbf{Coherent vector-meson production on a bound $\alpha$,
		      $A(e,e^\prime\alpha V)X$ with $V\in\{\phi,\,\mathrm{J/\psi}\}$.}
	      The principal observables are the $|t|$-differential cross sections,
	      expressed as ratios to the free-$^4$He cross sections measured in the
	      companion proposals~\cite{ALERT:Cprime}. Any shift in (or appearance of) a
	      diffractive minimum in the gluonic transition form factor is a direct
	      signature of medium modification of the gluon transverse distribution of
	      the bound $\alpha$.
\end{enumerate}
The same three observables form a coherent set: cross-section ratios and BSA
ratios remove many of the absolute normalization uncertainties and the
common-mode FSI corrections, leaving the bound-$\alpha$ partonic-structure
modification as the principal residual.

\subsection{Statistics and Systematics}


We will match the statistical precision of the ALERT program, specifically the measurements of coherent processes  on
$^4$He. By measuring cross section ratios and beam spin asymmetry ratios, many systematics effects will cancel.
Detector and DAQ efficiencies will need to be  monitored and ideally held constant through out the run 
to prevent large systematic corrections to the ratios. This systematic will be minimized through stable beam operations.

\subsection{Kinematics}
\label{sec:kinematics}

The kinematic coverage of the coherent measurements proposed here is
identical to that of the RG-L $E_{\mathrm{beam}}=10.6$~GeV running, in
which ALERT tags the recoiling $\alpha$ produced in coherent DVCS on
$^4$He, $^4\mathrm{He}(e,e^\prime\alpha\,\gamma)$, and coherent $\phi$
production on $^4$He, $^4\mathrm{He}(e,e^\prime\alpha\,\phi)$ with
$\phi\to K^+K^-$. Forming the bound-$\alpha$ observables of
Sec.~\ref{sec:proposed} as ratios to the free-$^4$He observables already
in analysis from RG-L therefore requires no additional kinematic
extrapolation. A separate 1-pass $E_{\mathrm{beam}}=2.2$~GeV
configuration is foreseen for the quasi-elastic $(e,e^\prime\alpha)$
channel; its dedicated kinematic settings will be detailed in the full
proposal. Quasi-elastic knockout kinematics are slightly more favorable 
for recoil detection because at the same
momentum transfer the recoil angle is slightly larger  
than for the DVCS or DVMP processes. 

\paragraph{Recoil-$\alpha$ acceptance in ALERT.}
The recoiling $\alpha$ from coherent DVCS and coherent $\phi$
production is detected in ALERT, installed in place of the CLAS12
silicon vertex tracker and operated inside the $5$~T solenoid. From the
combined AHDC+ATOF acceptance established during RG-L
commissioning~\cite{Armstrong:2017zcm} the recoil $\alpha$ is
reconstructed over the polar range
$\theta_{\alpha}\in[40^\circ,\,140^\circ]$ with full azimuthal
coverage, and above the momentum threshold, $ p_{\alpha}  > 240~\mathrm{MeV}/c$
corresponding to a kinetic-energy threshold of $T_{\alpha}\sim
8\text{--}10$~MeV (set by the AHDC/ATOF detection threshold). These 
recoils are well matched to the kinematics of coherent DVCS and DVMP,
where the $\alpha$ carries the small momentum transfer $|t|$. Particle
identification combines $dE/dx$ in the AHDC with time-of-flight in the
ATOF; both momentum and energy-loss information are required to
separate $^4$He from $d$, which share the same $m/q$.

\paragraph{Coherent kinematics on $^4$He.}
For coherent DVCS, $^4\mathrm{He}(e,e^\prime\alpha\,\gamma)$, and
coherent $\phi$ production, $^4\mathrm{He}(e,e^\prime\alpha\,\phi)$,
exclusivity is enforced by the missing-mass and missing-momentum
resolution of CLAS12  and ALERT, isolating the coherent peak at $|t|\sim 0$.
The recoil-$\alpha$ momentum threshold defined above translates the lowest
$|t|$  accessible with ALERT,  $|t| >  \approx 0.05~\mathrm{GeV}^2$, 
which covers the diffractive peak and extends past the region where the
first minimum occurs in the elastic  charge form  form factor. 
The full $\phi\in[0,2\pi)$ azimuthal
coverage of the recoil $\alpha$ is essential for the BSA extraction in
bins of the leptonic--hadronic plane angle.

\begin{figure}[h!]
\centering
\includegraphics[width=0.95\textwidth]{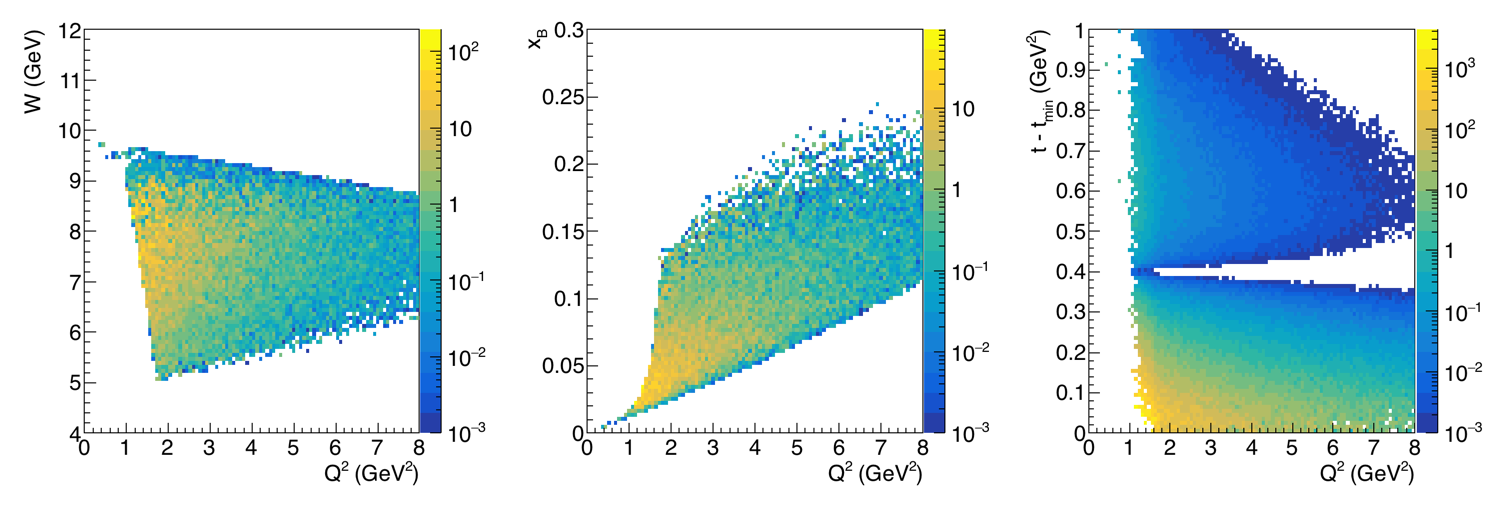}
\caption{Representative ALERT kinematic coverage for coherent $\phi$
electroproduction on $^4$He, $^4\mathrm{He}(e,e^\prime\alpha\,\phi)$,
at $E_{\mathrm{beam}}=10.6$~GeV with the $\alpha$ recoil detected in
ALERT. The accessible region in $Q^2$, $x_B$, and $|t|$ is set by the
CLAS12 forward acceptance for the scattered electron and $K^+K^-$
decay products combined with the recoil-$\alpha$ momentum and
angular acceptance of ALERT. Coherent DVCS on $^4$He
shares this recoil tag and populates the same $(x_B,|t|)$ region over
a comparable $Q^2$ range.}
\label{fig:phi_kinAcc}
\end{figure}

The combined $(Q^2,\,x_B,\,|t|)$ phase space accessible with the ALERT
recoil tag is illustrated in Fig.~\ref{fig:phi_kinAcc} for coherent
$\phi$ electroproduction on $^4$He. The figure is shown as a
representative example of the ALERT kinematic coverage; coherent DVCS
on $^4$He populates the same $(x_B,\,|t|)$ region with a comparable
(and somewhat broader) $Q^2$ reach, since both channels share the same
recoil-$\alpha$ acceptance and exclusivity selection.

\section{Estimated Beamtime Request}

Table~\ref{tab:beamtime} shows an estimated beam time request for a full proposal which matches the RG-L statistics 
for the coherent DVCS and DVMP channels. We also include some 1-pass running for the quasi-elastic knockout 
measurements. Rather than reusing RG-L helium target production data, we include it here to avoid potentially 
large and unknown systematic uncertainties entering the ratios. 

\begin{table}[h!]
	\centering
	\begin{tabular}{cccccc}
		Beam       & Beam    & Beam         & Target   & Material    & Beamtime   \\
		Energy     & Current & Requirements & Material & Thickness   & (days)     \\
		(GeV)      & (uA)    &              &          & (mg/cm$^2$) &            \\
		2.2        & 0.5     & Unpolarized  & Helium   & 27.1        & 1          \\
		\hline
		2.2        & 0.5     & Unpolarized  & Helium   & 27.1        & 1          \\
		2.2        & 0.5     & Unpolarized  & Oxygen   & 216         & 1          \\
		2.2        & 0.5     & Unpolarized  & Neon     & 136         & 1          \\
		2.2        & 0.5     & Unpolarized  & Ethane   & 203.1       & 1          \\
		\hline
		10.6       & 0.5     & Polarized    & Helium   & 27.1        & 20         \\
		10.6       & 0.062   & Polarized    & Oxygen   & 216         & 20         \\
		10.6       & 0.099   & Polarized    & Neon     & 136         & 20         \\
		10.6       & 0.083   & Polarized    & Ethane   & 203.1       & 20         \\
		           &         &              &          &             &            \\
		\hline
		Total Time &         &              &          &             & 85  \\
	\end{tabular}
	\caption{Estimated Beam time request for $\alpha$-knockout measurements. Low-energy
		runs (2.2~GeV) provide calibration data; high-energy runs (10.6~GeV)
		target the DVCS/DVMP channels.}
	\label{tab:beamtime}
\end{table}


\section{Summary}


We propose the first comprehensive program of hard exclusive
$\alpha$-knockout measurements from light $\alpha$-conjugate nuclei. The
combination of quasi-elastic $(e,e^\prime\alpha)$ knockout, coherent DVCS on
a bound $\alpha$, and coherent vector-meson production on a bound $\alpha$,
measured on $^4$He, $^{12}$C, $^{16}$O, and $^{20}$Ne with the same
apparatus and analysis chain, provides a direct partonic-level probe of the
quark and gluon distributions of an $\alpha$ embedded in heavier nuclei. The
observables are cross-section and beam-spin-asymmetry ratios to free $^4$He,
which are largely insensitive to the absolute normalization and to
common-mode FSI corrections. The pattern of these ratios across the
$\alpha$-conjugate chain---in which the ground-state $\alpha$-clustering
strength varies from $\sim 6\%$ in $^{12}$C to $\sim 70\%$ in
$^{20}$Ne---directly tests whether QCD-level modifications of the bound
$\alpha$ track its cluster environment.

The measurement uses CLAS12 with the ALERT detector and high-pressure gas
straw targets in Hall~B. ALERT was successfully commissioned and
operated during RG-L (April--August 2025), providing for the first time the
low-momentum, light-ion recoil tagging required for these exclusive
channels. The proposed program is a natural extension of the ongoing ALERT
physics program on $^4$He and of the conditionally approved coherent-J/$\psi$
measurement on $^4$He~\cite{ALERT:Cprime}. We estimate a beam time request totaling  roughly  \textbf{85 PAC days}.

\newpage

%
%
%
%
\bibliographystyle{unsrt}
\bibliography{references}

\newpage

\appendix

\section{Survey of $(e,e^\prime\alpha)$ and coherent-$\alpha$
electron-scattering experiments}
\label{app:survey}

This appendix collects, to the best of our knowledge, all published
electron-scattering experiments in which a final-state $\alpha$ particle
($^4$He) was directly detected. Two classes of measurement are
distinguished: (i) quasi-elastic $(e,e^\prime\alpha)$ knockout, in which
the $\alpha$ is liberated from a heavier target nucleus, and (ii)
coherent processes on $^4$He, in which the recoiling $\alpha$ is the
intact target nucleus. We list here only experiments performed with
electron beams; companion measurements with hadronic probes (such as
$(p,p'\alpha)$, $(\alpha,2\alpha)$, or $(^{6}\mathrm{Li},d)$) are
extensively catalogued in the cluster-physics
literature~\cite{Freer:2017gip} and are not reproduced
here. Theoretical analyses (e.g.\ DWIA frameworks) are referenced where
they form an inseparable part of the experimental program but are not
themselves listed as experiments.

\subsection*{Quasi-elastic $(e,e^\prime\alpha)$ measurements}

\begin{table}[h!]
	\centering\small
	\renewcommand{\arraystretch}{1.25}
	\begin{tabular}{lllllc}
		\hline
		Year & Reaction & Facility & Lead author & Citation & Ref. \\
		\hline
		1974 & $^{6}$Li$(e,e^\prime\alpha)$d                & Saclay LINAC      & Genin       & Phys.\ Lett.\ B \textbf{52}, 46     & \cite{Genin:1974zz}    \\
		1974 & $^{6}$Li$(e,e^\prime\alpha)$d  (theory)      & ---               & Toyama      & Phys.\ Lett.\ B \textbf{51}, 217    & \cite{Toyama:1974zz}    \\
		1986 & $^{6}$Li$(e,e^\prime d)^4$He$^{\,(*)}$       & NIKHEF MEA        & Ent         & Phys.\ Rev.\ Lett.\ \textbf{57}, 2367 & \cite{Ent:1986rc}      \\
		1991 & $^{6}$Li$(e,e^\prime\alpha)$                 & NIKHEF AmPS       & Mitchell    & Phys.\ Rev.\ C \textbf{44}, 2002    & \cite{Mitchell:1991sb} \\
		1995 & $^{12}$C$(e,e^\prime\alpha)^8$Be             & Mainz / NIKHEF    & DeAngelis   & Phys.\ Rev.\ C \textbf{52}, 61      & \cite{DeAngelis:1995zz} \\
		2001 & $^{16}$O$(e,e^\prime\alpha)^{12}$C           & NIKHEF AmPS       & De Meyer    & Phys.\ Lett.\ B \textbf{513}, 258   & \cite{DeMeyer:2001uv}  \\
		\hline
	\end{tabular}
	\caption{Published electron-scattering measurements with a directly
		detected final-state $\alpha$. The 1986 NIKHEF entry is the
		complementary $(e,e^\prime d)^4$He channel on the same $^6$Li
		ground state, retained here because it provides the
		$\alpha$\textendash$d$ momentum distribution and uses the same
		detector configuration. The 1974 Toyama--Sakamoto entry is the
		companion microscopic calculation that established the PWIA
		analysis of the Saclay $^{6}$Li data.}
	\label{tab:eeAlphaExperiments}
\end{table}

\paragraph{Saclay (1974).} The first $(e,e^\prime\alpha)$ measurement was
performed by Genin {\it et al.} on $^6$Li at $E_e = 520$~MeV using the
Saclay 600~MeV LINAC~\cite{Genin:1974zz}. Coincident detection of the
scattered electron and the recoiling $\alpha$ from
$^6\mathrm{Li}\rightarrow d + \alpha$ established the quasi-free
$\alpha$-knockout mechanism and provided the first electron-scattering
measurement of an $\alpha$-cluster momentum distribution. The companion
PWIA/microscopic analysis was published simultaneously by Toyama and
Sakamoto~\cite{Toyama:1974zz}.

\paragraph{NIKHEF-K (1986--2001).} The NIKHEF program at the MEA
linac and later the AmPS storage ring produced the most complete body
of $(e,e^\prime\alpha)$ data to date. The Ent {\it et al.}\ 1986
measurement of $^6\mathrm{Li}(e,e^\prime d)^4\mathrm{He}$~\cite{Ent:1986rc}
extracted the $\alpha$\textendash$d$ momentum distribution to
$p_m\simeq 270$~MeV/$c$ and confirmed the dominance of the
$\alpha$\textendash$d$ component in the $^6$Li ground state. Mitchell
{\it et al.}\ \cite{Mitchell:1991sb} subsequently measured
$^6\mathrm{Li}(e,e^\prime\alpha)$ in parallel kinematics, isolating the
quasi-elastic mechanism through the $Q^2$ dependence of the cross
section. The flagship measurement of the program was the
$^{16}\mathrm{O}(e,e^\prime\alpha)^{12}\mathrm{C}$ experiment of De Meyer
{\it et al.}~\cite{DeMeyer:2001uv} at AmPS, which achieved
$\sim 700$~keV missing-energy resolution\textemdash a factor-of-two
improvement over previous $(e,e^\prime\alpha)$ data\textemdash and
mapped the missing-momentum distribution out to $\sim 300$~MeV/$c$.
The data confirm the quasi-elastic $\alpha$-knockout mechanism but
overshoot DWIA calculations using contemporary cluster-model
$\alpha$-cluster wave functions by a factor of $3$\textendash$4$,
indicating sizeable model dependence in the extracted $\alpha$
spectroscopic factors.

\paragraph{Mainz MAMI (1995).} The DeAngelis
{\it et al.}\ measurement of $^{12}\mathrm{C}(e,e^\prime\alpha)$ was
performed at the 300~MeV electron accelerator at
Mainz~\cite{DeAngelis:1995zz} and analyzed at NIKHEF. Charged-particle
emission to both the ground state and the $2^+$ state of $^8$Be was
measured at four values of the momentum transfer
($q = 0.24$\textendash$0.61$~fm$^{-1}$), and a multipole decomposition
established the $E2$ dominance of the $\alpha$ emission, with
non-negligible $E0$ and $E3$ contributions. The strengths were
compared with $(p,p'\alpha)$ and $(\alpha,\alpha',\alpha)$ data and
provide a multipole baseline against which the proposed CLAS12+ALERT
$(e,e^\prime\alpha)$ measurements at $10.6$~GeV may be benchmarked at
low $|t|$ and forward kinematics.

\paragraph{Status.} Beyond the entries in
Table~\ref{tab:eeAlphaExperiments} we are not aware of any other
peer-reviewed $(e,e^\prime\alpha)$ measurement on
$^7$Li, $^9$Be, $^{12}$C, $^{16}$O, $^{20}$Ne, $^{24}$Mg, or
$^{40}$Ca. In particular, no $(e,e^\prime\alpha)$ data exist on
$^{12}$C beyond the $(e,e^\prime\alpha)^8$Be channel of
Ref.~\cite{DeAngelis:1995zz}; no measurements of
$^{12}$C$(e,e^\prime\alpha)2\alpha$ (three-body breakup) and no
measurements on $^{20}$Ne have been reported. The proposed measurements
in this LOI would therefore constitute the first $(e,e^\prime\alpha)$
data ever taken on $^{20}$Ne and the first three-body
$(e,e^\prime\alpha)2\alpha$ data on $^{12}$C, in addition to extending
the $^{16}$O data of Ref.~\cite{DeMeyer:2001uv} by more than three
orders of magnitude in luminosity and into the deeply-virtual regime.

\subsection*{Coherent processes on $^4$He with detected recoil $\alpha$}

The coherent-$^4$He channels are not $(e,e^\prime\alpha)$ in the
knockout sense, but they share the experimental signature of an
intact recoiling $\alpha$ in the final state and are conceptually
the free-$\alpha$ baselines for the proposed bound-$\alpha$
measurements.

\begin{table}[h!]
	\centering\small
	\renewcommand{\arraystretch}{1.25}
	\begin{tabular}{lllllc}
		\hline
		Year & Reaction & Facility / detector & Lead author & Citation & Ref. \\
		\hline
		2014 & $^{4}$He$(e,e^\prime\,^4$He$)$ elastic              & JLab Hall A           & Camsonne   & Phys.\ Rev.\ Lett.\ \textbf{112}, 132503 & \cite{Camsonne:2014vfa} \\
		2017 & $^{4}$He$(e,e^\prime\,^4$He$\,\gamma)$ coherent DVCS & JLab CLAS / EG6      & Hattawy    & Phys.\ Rev.\ Lett.\ \textbf{119}, 202004 & \cite{CLAS:2021ovm}    \\
		2019 & $^{4}$He$(e,e^\prime p\,\gamma)$ incoherent DVCS    & JLab CLAS / EG6       & Hattawy    & Phys.\ Rev.\ Lett.\ \textbf{123}, 032502 & \cite{Hattawy:2018liu} \\
		\hline
	\end{tabular}
	\caption{Published electron-scattering experiments on $^4$He with
		direct detection of the recoil $^4$He (or, for the incoherent
		channel, the proton knocked out of $^4$He). The Camsonne {\it et
		al.}\ measurement extends the $^4$He charge form factor across
		its diffractive minimum; the Hattawy {\it et al.}\ measurements
		are the first exclusive coherent and incoherent DVCS data on a
		nucleus heavier than the deuteron, both performed with the
		BoNuS-style radial time-projection chamber installed in CLAS
		during EG6.}
	\label{tab:coherent4He}
\end{table}

To our knowledge, no peer-reviewed measurement of coherent
$\phi$ or $\mathrm{J}/\psi$ \emph{electroproduction} on $^4$He with a
detected recoil $\alpha$ exists; the closest companion datasets are
the LEPS coherent $\phi$-photoproduction measurement on
$^4$He~\cite{ALERT:Cprime} and the recently conditionally approved
JLab proposal C12-25-004 on coherent $\mathrm{J}/\psi$
electroproduction on $^4$He with CLAS12+ALERT~\cite{ALERT:Cprime},
which together motivate the present LOI.

\end{document}

%% file: authors.tex
%
%
%

%
%

\author{Whitney~R.~Armstrong$^{\dagger}$~\orcidlink{0000-0001-7359-6341}} 
\author{Ian Clo\"et~\orcidlink{0000-0003-4877-2862}} 
\author{Fernando A. Flor~\orcidlink{0000-0002-0194-1318}} 
\author{Sylvester~Joosten~\orcidlink{0000-0003-4947-877X}} 
\author{Bobae~Kim~\orcidlink{0000-0002-9539-6815}} 
\author{Minho~Kim~\orcidlink{0000-0002-3581-8342}} 
\author{Henry~T.~Klest~\orcidlink{0000-0003-4695-0223} } 
\author{Zein-Eddine~Meziani~\orcidlink{0000-0001-9450-2914}} 
\author{Chao~Peng~\orcidlink{0009-0002-3353-0510}} 
\author{Valerii~Klimenko~\orcidlink{0000-0003-4159-9833}} 
\author{Noémie~Pilleux~\orcidlink{0000-0003-1174-5684}} 
\author{Maria~Żurek~\orcidlink{0000-0003-4201-4921}} 
\author{Jacob~C.~Dulya~\orcidlink{0000-0002-5101-8580}} 
\author{Zhiwan~Xu~\orcidlink{0000-0003-3999-0585}} 
\affil{Argonne National Laboratory, Lemont, Illinois 60439, USA}

\author{Lamiaa~El~Fassi~\orcidlink{0000-0003-3647-3136}} 
\affil{Mississippi State University, Mississippi State, Mississippi 39762, USA}

\author{Penn~Smith} 
\author{Nazmus~S.~Ifat} 
\author{Sangbaek~Lee~\orcidlink{0000-0001-6038-0736}} 
\author{Suman~Shrestha} 
\author{Nikos~Sparveris~\orcidlink{0000-0002-6278-1178}} 
\affil{Temple University, Philadelphia, PA 19122, USA}

\author{Fatiha~Benmokhtar~\orcidlink{0000-0003-3670-4045}} 
\affil{Duquesne University, Pittsburgh, PA 152082, USA}

\author{Christine~Ploen$^{ }$~\orcidlink{0000-0002-6237-3692}} 
\affil{Old Dominion University, Norfolk, VA 23529, USA }

\author{Raphaël Dupré~\orcidlink{0000-0001-7935-8221}} 
\affil{Université Paris-Saclay, CNRS, IJCLab, 91405, Orsay, France.}

%
\author{Jennifer~Rittenhouse~West$^\ddagger$~\orcidlink{0000-0002-8725-2184}} 
\affil{INFN Section Turin, Via Pietro Giuria 1, 10125 Turin, Italy}

\author{Stanley~J.~Brodsky$^\ddagger$~\orcidlink{0000-0001-8786-3172}} 
\affil{SLAC National Accelerator Laboratory, Stanford University, Stanford, California 94309, USA}


\date{A CLAS Collaboration Letter of Intent\\
  \vspace{2mm}
\small{$\dagger$: Contact-spokesperson (\href{mailto://warmstrong@anl.gov}{warmstrong@anl.gov}), $\ddagger$: Theory Support}}
     
%